\documentclass[12pt]{article}

\usepackage{scicite}

\usepackage{times}
\usepackage{graphicx}
\usepackage{graphics}
\usepackage{amsmath}
\usepackage{color}

\newenvironment{sciabstract}{%
\begin{quote} \bf}
{\end{quote}}

\newcounter{lastnote}
\newenvironment{scilastnote}{%
\setcounter{lastnote}{\value{enumiv}}%
\addtocounter{lastnote}{+1}%
\begin{list}%
{\arabic{lastnote}.}
{\setlength{\leftmargin}{.22in}}
{\setlength{\labelsep}{.5em}}}
{\end{list}}

\title{Probing Interactions between Ultracold Fermions}

\author
{G. K. Campbell$^1$, M. M. Boyd$^1$, J. W. Thomsen$^1$,  M. J.
Martin$^1$, \\S. Blatt$^1$,  M. D. Swallows$^1$, T. L.
Nicholson$^1$, T. Fortier$^2$,\\ C. W. Oates$^2$, S. A. Diddams$^2$,
N. D. Lemke$^2$,
P. Naidon$^3$, \\P. Julienne$^3$, Jun Ye$^{1\ast}$ \& A. D. Ludlow$^1$\\
\\
\normalsize{$^{1}$JILA, National Institute of Standards and
Technology and University of Colorado} \\
\normalsize{Department of Physics, University of
Colorado, Boulder, Colorado 80309-0440, USA}\\
\normalsize{$^{2}$Time and Frequency Division, National Institute of Standards and Technology,}\\
\normalsize{Boulder, CO 80302, USA}\\
\normalsize{$^{3}$ Atomic Physics Division and Joint Quantum
Institute,}\\ \normalsize{National Institute of Standards and Technology,} \\
\normalsize{100 Bureau Drive Stop 8423, Gaithersburg, Maryland
20899-8423}
\\
\normalsize{$^\ast$To whom correspondence should be addressed;
E-mail:  junye@jilau1.colorado.edu.}
}

\date{}

\begin{document}


\baselineskip24pt


\maketitle


\begin{sciabstract}
At ultracold temperatures, the Pauli exclusion principle suppresses
collisions between identical fermions. This has motivated the
development of atomic clocks using fermionic isotopes. However, by
probing an optical clock transition with thousands of
lattice-confined, ultracold fermionic Sr atoms, we have observed
density-dependent collisional frequency shifts. These collision
effects have been measured systematically and are supported by a
theoretical description attributing them to inhomogeneities in the
probe excitation process that render the atoms distinguishable. This
work has also yielded insights for zeroing the clock density shift.
\end{sciabstract}

Quantum statistics play a critical role in shaping interactions
between matter. This is apparent in the markedly different behavior
of Bose-Einstein condensates\cite{Cornell02,Ketterle02} and
degenerate Fermi gases of ultracold atoms\cite{Demarco99b}.  The
quantum statistics of atoms can thus be a key factor in
the choice of an atomic system for a given experiment. Such is the
case for atoms at the heart of an atomic clock. Simultaneous
interrogation of many atoms is favorable for achieving high
measurement precision. However, when atoms interact with each other,
their internal energy states can be perturbed, leading to frequency
shifts of the clock transition\cite{Gibble93,Sortais01}. The use of
identical fermions was prescribed to allow many atoms to strengthen
the signal without such density-dependent collision
shifts\cite{Gibble95}. Previous experiments seemed to confirm this
fact for both single-component\cite{Gupta03} and two-component
fermion mixtures\cite{Zwierlein03}.

However, by probing an optical clock transition with thousands of
fermionic Sr atoms confined in a one-dimensional optical lattice, we
clearly observe density-related frequency shifts at a fractional
precision of 1$\times10^{-16}$. When the light-atom interaction
introduces a small degree of inhomogeneous excitation, previously
indistinguishable fermions become slightly distinguishable. This
effect causes a time-dependent variation of the two-particle
correlation function, giving rise to an apparent mean-field energy.
The resulting collision effects have been measured systematically as
a function of temperature, excitation probability, and interaction
inhomogeneity. These observations are supported by a theoretical
description of fermionic interactions including the effect of the
measurement process.

The latest generation of optical atomic clocks such as those based
on the $^1$S$_0-^3$P$_0$ transition in fermionic $^{87}$Sr currently
offers the highest measurement precision, useful for measuring possible
atomic interactions\cite{Ludlow08,Campbell08}. In an ultracold
dilute gas with a mean field energy, a narrow clock transition will
experience a density-dependent frequency shift\cite{Leo01,Harber02}
given by $h\Delta\nu=(4\pi\hbar^2 G^{(2)}\rho a)/m$. Here $m$ is the
atomic mass, $\rho$ is the density of the atomic sample, $a$ is the
$s$-wave scattering length characterizing the atomic interaction,
and $h=2\pi\hbar$ is Planck's constant. $G^{(2)}$ is the two-atom
correlation function at zero distance, which summarizes the quantum
statistics of colliding bodies. For example, $G^{(2)}=0$ for
identical fermions and $G^{(2)}=2$ for identical bosons in a thermal
gas. The Fermi suppression arises from the Pauli exclusion
principle, which prohibits even-partial-wave collisions between
indistinguishable fermions. At ultracold temperatures partial waves
higher than $s$-wave are frozen out\cite{Demarco99}. For atoms
excited in our two-level clock system, three possible $s$-wave
interactions exist: those between two $^1$S$_0$ ground-state
($|g\rangle$) atoms, those between two $^3$P$_0$ excited-state
($|e\rangle$) atoms, and those between a $|g\rangle$ atom and a
$|e\rangle$ atom. Including all possible interactions, the
collisional frequency shift at ultracold temperatures is given by
Eq. 1 \cite{Zwierlein03,Harber02,Leo01}:
\begin{equation}
\Delta\nu_{ge}=\frac{2\hbar}{m}(G^{(2)}_{ge}a_{ge}(\rho_g-\rho_e)+G^{(2)}_{ee}a_{ee}\rho_e-G^{(2)}_{gg}a_{gg}\rho_g),
\end{equation}
where $a_{ij}$ is the $s$-wave scattering length for collisions
between atoms in state $i$ and $j$, and $\rho_i$ is the density of
atoms in state $i$. Since indistinguishable fermions do not collide,
$G^{(2)}_{gg}$ = $G^{(2)}_{ee}$ = 0. Fermions in different internal
states are distinguishable, and for a completely incoherent mixture
of the two states, $G^{(2)}_{ge}=1$. However, if the two-state
mixture is prepared by a uniform, coherent excitation of
ground-state atoms, then the fermions evolve indistinguishably and
$G^{(2)}_{ge}$ = 0\cite{Zwierlein03}. In this case,
$\Delta\nu_{ge}=0$.

Two possibilities exist for $\Delta\nu_{ge}$ to deviate from zero.
First, the $p$-wave contribution may not be negligible. However, for
ultracold atoms confined in a well-characterized optical trap, we
show experimental evidences and theoretical calculations that
conclude that $p$-wave collisions make no noticeable contribution to
the observed clock frequency shift. Second, it is imperative to
consider the entire interaction, including the measurement process,
when exploring the question of whether fermions collide. In fact,
the measurement process, such as probing a clock transition, may
strongly influence the time-dependent variations of $G^{(2)}$. We
show here that an inhomogeneous interaction between light and atoms
leads to the loss of indistinguishability of the fermions, thus
making $0<G^{(2)}<1$.

Although a uniform, coherent excitation of identical fermions
maintains $G^{(2)}=0$, and no $s$-wave collisions occur, if a small
non uniformity in the excitation process arises, the atoms are no
longer completely identical, and $G^{(2)}>0$. The value of $G^{(2)}$
will depend on the degree of excitation inhomogeneity. This
measurement-induced dynamic variation of quantum statistics leads
directly to a change of the mean-field energy within the ultracold
gas, resulting in a nonzero $\Delta\nu_{ge}$. It is interesting to
contrast the present work with previous results observed using an
ultracold gas of fermionic ${^6}$Li, where the insensitivity of a
radio-frequency transition to collisional shifts was
demonstrated\cite{Gupta03,Zwierlein03}. It was shown that the
fermionic insensitivity to collisional shifts was maintained even
when a pure superposition state of the two level system had
decohered. This decoherence would have allowed interactions, but
when a uniform rf probing field reintroduced coherence to the atoms
in a homogeneous manner, $G^{(2)}$ again became zero, giving no
collisional shifts within the measurement precision\cite{MITnote}.
>From the current experiment it is clear that any non-identical evolutions
during the interrogation process lead to the breakdown of Fermi
suppression; this experiment is sensitive to very small
inhomogeneities because of the high measurement precision.

An intuitive understanding emerges from considering two sample atoms
in a pseudo spin-1/2 system with ground $|g\rangle$ and excited
$|e\rangle$ states. Before applying the spectroscopy pulse, the
atomic system is in a pure, polarized spin state with
$|\psi_1\rangle=|\psi_2\rangle=|g\rangle$. The effect of the pulse
is to perform a rotation on the Bloch sphere, as shown in the inset
of Fig. 1B. For a coherent, homogeneous excitation, the wavefunction
of the system becomes a coherent superposition
$|\psi_1\rangle=|\psi_2\rangle=\alpha|g\rangle+\beta|e\rangle$. The
wavefunctions of both atoms are identical, $G^{(2)}_{12}$ = 0, and
collisions cannot occur. An inhomogenous spectroscopic excitation, such as that caused by varying Rabi frequencies for different atoms,
results in slightly different rotations on the Bloch sphere for the
two atoms (Fig. 1B inset). Hence we have
$|\psi_1\rangle=\alpha|g\rangle+\beta|e\rangle$ and
$|\psi_2\rangle=\gamma|g\rangle+\delta|e\rangle$. The fermions are
distinguishable and 0 $< G^{(2)}_{12}< $1. The value of
$G^{(2)}_{12}$ depends on the amount of inhomogeneity, and its time
variation can be explicitly calculated from the anti-symmetrized
overlap of the two wavefunctions (details in supporting
text\cite{SOM}):
\begin{equation}
G^{(2)}_{12}(\alpha(t),\beta(t),\gamma(t),\delta(t))=1-|\alpha(t)\gamma^*(t)+\beta(t)\delta^*(t)|^2.
\end{equation}
The resulting collision shift from Eq. 1 is then
\begin{equation}
\Delta\nu(t)=\frac{2\hbar
a_{ge}}{m}G^{(2)}_{12}(\alpha,\beta,\gamma,\delta)(\rho_g-\rho_e).
\end{equation}

Before proceeding with experimental results, we first summarize the
system under study\cite{SOM}. In the $^{87}$Sr optical clock, atoms
are trapped in a one-dimensional optical standing-wave potential (1D
optical lattice). Longitudinally the atoms are confined tightly,
with an oscillation frequency $\nu_z$ $\sim$80 kHz. At temperature
$T$ = 1 $\mu$K, $\sim$98 \% of the atoms occupy the ground-state of
the trap ($\bar{n}_z$ = 0.02). The laser probing the clock
transition propagates along the lattice axis, and spectroscopy is
performed in the Lamb-Dicke regime. In the transverse plane the
confinement is much weaker, with an oscillation frequency
$\nu_x=\nu_y\sim$450 Hz, and atoms occupy a large number of motional
states ($\bar{n}_x$ = $\bar{n}_y$ = 46). Typically,
$\sim$2$\times$10$^{3}$ atoms are trapped in the optical lattice,
resulting in 30 atoms per lattice site with a density of
2$\times$10$^{11}$ cm$^{-3}$\cite{SOM}. The optical lattice is
nearly vertically oriented and is operated at the so called ``magic
wavelength'' of $\lambda_L$ $\sim$813.429 nm\cite{Ye08}, where the
ac Stark shifts of the $^1$S$_0$ and $^3$P$_0$ states are identical.

With a perfect alignment of the probe
laser along the strong confinement axis, assuming cylindrical
symmetry, a residual angular spread between the probe
and lattice $\vec{k}$ remains due to the finite size of the lattice beam\cite{Pritchard88}. However, an even larger effect occurs if
the symmetry is broken due to either aberrations in the
beam profile or angular misalignment ($\Delta\theta$) between
the lattice and the probe beam. For our trap parameters, we estimate an effective $\Delta\theta\approx$ 10 mrad (Fig.1A inset). The
residual wave-vector projected on the transverse plane leads to
slightly different excitation Rabi frequencies $\Omega_{\vec{n}}$
for atoms in different ($n_x$, $n_y$)
states\cite{SOM,Wineland79,Akatsuka08}. For a given $T$, the occupation of a transverse motional state
$n_{x,y}$ is given by the normalized Maxwell-Boltzmann
distribution. The inhomogeneity in the
Rabi frequencies is thus affected by both $T$ and $\Delta\theta$.

To calculate the density shift, we return to our two-atom model.
Each atom has a slightly different $\Omega_{\vec{n}}$. For the
entire atomic ensemble, we can define an average Rabi frequency
$\overline{\Omega}$ and its RMS spread $\Delta\Omega$. To
approximate the average density shift, we set $\Omega_1 =
\overline{\Omega}+\Delta\Omega$ and $\Omega_2 =
\overline{\Omega}-\Delta\Omega$ for our two-atom model. Thus the
time-dependent quantities, $\alpha$, $\beta$, $\gamma$, $\delta$
as defined in Eq. 2 are parameterized by $\overline{\Omega}$ and
$\Delta\Omega$\cite{SOM}. At a time $t$ during the spectroscopy pulse,
the atoms experience an ensemble-averaged shift:
\begin{equation}
\Delta\nu(t)=\frac{2\hbar
a_{ge}}{m}G^{(2)}_{12}(\overline{\Omega}+\Delta\Omega,\overline{\Omega}-\Delta\Omega)(\rho_g-\rho_e).
\end{equation}
This shift evolves during the spectroscopy pulse, and for the final
density shift we time average $\Delta\nu(t)$ over the total pulse
length $t_F$. This approximation is valid in the limit that the
change in $\Omega$ due to atomic interactions is much less than
$\Delta\Omega$.  A more rigorous calculation using the optical Bloch
equations including atomic interactions has also been made. Using
our typical trap parameters we find the two-atom approximation is
valid to within 5 \%. The time-dependent Rabi-oscillation is only
slightly affected by atomic interactions, however the effect on the
final clock shift is obvious.

For inhomogeneity-induced collision shifts, $t_F$ is important.
Atoms in close proximity to each other tend to have similar Rabi
frequencies, whereas atoms located far apart are more likely to
experience different excitations (and hence be distinguishable). If
$t_F\nu_{x,y}\ll1$, the atoms are effectively frozen in place, and
will experience no density shift. However, if $t_F\nu_{x,y}>1$,
atoms initially located far apart have time to interact. For the clock
experiment requiring high spectral resolution, we have $t_F$ = 80 ms and $1/\nu_{x,y}$ = 2.2 ms, so
collisions will occur.

To systematically study these effects, we implemented controlled
variations of both $T$ and $\Delta\theta$. To vary $T$, we perform
cooling (heating) of the lattice-confined atoms in three dimensions:
Doppler cooling (heating) along the transverse direction and
sideband cooling (heating) along the longitudinal axis. Simultaneous
with the sideband cooling (heating), the atoms are spin-polarized by
optical pumping in a weak magnetic ($B$) bias field. Atoms are
polarized into either the $m_F$ = +9/2 or $m_F$ = -9/2 Zeeman
states. The $^1$S$_0$ $-^3$P$_0$ clock transition, which is
predicted to have a natural linewidth of $\sim$ 1
mHz\cite{boyd07b,Santra04,Porsev04}, is interrogated using a
cavity-stabilized diode laser at 698 nm with a linewidth below 1
Hz\cite{Ludlow07}. Spectroscopy is performed in the Lamb-Dicke
regime and in the resolved sideband limit\cite{Leibfried03}. To
ensure that the polarized spin state is well resolved from other
$m_F$ levels, spectroscopy is performed under $B$ $\sim$250 mG,
leading to a separation of 250 Hz between the $m_F = \pm$9/2 states.
A spectroscopy pulse length of $t_F$ = 80 ms results in a
Fourier-limited linewidth of $\sim$10 Hz.

After the spectroscopy pulse is applied, atoms remaining in
$|g\rangle$ are counted by measuring fluorescence on the strong
$^1$S$_0$$- ^1$P$_1$ transition. Atoms transferred to $|e\rangle$
are then pumped back to $|g\rangle$ via the intermediate
(5s6s)$^3$S$_1$ states and are also counted. Combining these two
measurements gives us a normalized excitation fraction
$\rho_e/(\rho_e + \rho_g)$. The atomic temperature is determined
using both sideband spectroscopy\cite{Thomsen08,SOM} and
time-of-flight analysis. In Fig. 1A, sample spectra are shown for
two different values of $T$. Once $T$ is measured, the degree of
inhomogeneity is determined by fitting the decaying Rabi
oscillations for the ensemble. In Fig. 1B, the Rabi oscillation at
$T$ = 3 $\mu$K (squares) clearly shows faster dephasing than that of
$T$ = 1 $\mu$K (circles), indicating a larger degree of
inhomogeneity.

Density-dependent frequency shifts of the $^{87}$Sr clock transition
are measured using a remotely located calcium optical standard at
NIST\cite{Ludlow08} as a stable frequency reference, which is linked
to JILA via a phase-coherent fiber network\cite{foreman07b}. This
direct optical frequency measurement between two optical standards
allows fractional measurement precision of a few times $10^{-16}$
after hundreds of seconds of averaging.  To measure the clock center
frequency, the spectroscopy pulse is first applied to atoms
optically pumped to the $m_F$ = +9/2 state. In the next cycle, atoms
polarized to the $m_F$ = -9/2 state are used. The center frequency
is then determined by the average of both resonances. The
density-dependent frequency shift is determined using an interleaved
scheme, where the density of the atomic ensemble is varied every 100
s. The density is varied by a factor of two. Pairs of such
data are then used to measure a frequency shift, and many pairs are
averaged to decrease the statistical uncertainty. Typically, we lock
the clock laser near the full-width at half-maximum of each
resonance, however the location of the lock points is varied to
select the desired excitation fraction.

Spectroscopy is performed using two different experimental
procedures. In the first, we probe the clock transition from
$|g\rangle$ to $|e\rangle$ (Fig. 2 inset). The intensity of the
probe is set to produce a $\pi$-pulse on resonance. This direct
scheme could suffer from imperfect polarization of the atomic
sample, and spectator atoms could be left in other $m_F$ levels.
This scenario could potentially lead to density-dependent shifts due
to collisions between different $m_F$ states that are not suppressed
by the Fermi statistics. The second scheme minimizes this effect by
probing $|e\rangle$ to $|g\rangle$ (Fig. 2). Here we apply a strong pulse to first transfer the population from $|g\rangle$ to
$|e\rangle$. The pulse power broadens the transition in order to decrease the
sensitivity of population transfer to probe laser frequency, and transfers
$\sim$50 \% of the population to $|e\rangle$. This first pulse is
resonant with atoms in one of the $m_F=\pm$ 9/2 states, hence atoms
left in other $m_F$ states due to imperfect polarization are not
transferred. Subsequently, all atoms remaining in $|g\rangle$ are
removed from the lattice with a pulse of light resonant with the
strong $^1$S$_0$$- ^1$P$_1$ transition, without affecting the
temperature of the atoms in $|e\rangle$. This is confirmed with
sideband spectroscopy\cite{SOM}. Finally, the clock transition of
$|e\rangle$ to $|g\rangle$ is probed with the usual 80 ms
$\pi$-pulse. In both experimental procedures, we measure populations
in $|e\rangle$ and $|g\rangle$ to determine the normalized
excitation fraction.

Figure 2 summarizes the measured density-dependent frequency shift
as a function of the normalized ground-state fraction for two
different values of $T$, 1 $\mu$K (squares) and 3 $\mu$K (circles).
The data indicate a clear trend that the density shift decreases
under a more homogeneous excitation. The solid lines are the
expected shifts calculated from the two-atom model. For clock
operation, it is important to note that near 50 \% excitation
fraction, for both $T$, the shift goes through zero.

Of course, as we change $T$, we vary both the excitation
inhomogeneity and the $p$-wave contribution. To estimate the
magnitude of $p$-wave collisions, we note that the van der Waals
potential for all three interaction types ($gg$, $ee$, or $eg$) has
been theoretically calculated\cite{Porsev02,Santra04,JulienneSP}, and the
$p$-wave centrifugal barrier is expected to be greater than 25
$\mu$K. At $T$ $\sim$1 $\mu$K, $ka\ll 1$, where
$k=2\pi/\lambda_{T}$. $\lambda_{T}=h/\sqrt{2\pi m k_BT}$ is the
thermal deBroglie wavelength, and $k_B$ is the Boltzmann constant.
Under these conditions, the ratio of $p$-wave to $s$-wave phase
shift is $(b k)^2 b/a$, where $b$ is the $p$-wave scattering length.
For $gg$ interactions, the $s$-wave scattering length has been
measured\cite{Martinez08} for $^{88}$Sr, and mass scaling gives
$a_{gg}$= 96.2(1)$a_0$ for $^{87}$Sr, where $a_0$ is the Bohr
radius. Combined with the van der Waals potential, the $p$-wave
phase shift can be determined from the Schr\"{o}dinger equation. For
$^1$S$_0$, $b_{gg}$ = -76 $a_0$, and for $T$ = 1 $\mu$K, $|(b_{gg}
k)^2 b_{gg}/a_{gg}|\approx 0.01$. Thus, $p$-wave collisions for $gg$
are suppressed by over two orders of magnitude and are negligibly
small. Although the $s$-wave scattering lengths $a_{ee}$ and
$a_{ge}$ have not yet been measured and thus cannot directly
constrain the values of $b_{ee}$ and $b_{eg}$, calculations based on
a theoretical potential predict that these $p$-wave collisions are
similarly suppressed relative to $s$-wave collisions. An exception
would be a $p$-wave shape resonance\cite{Demarco99}; however, this
would occur only for a very small range of possible $a_{ee}$ and
$a_{ge}$, and the effect would be reduced by thermal averaging.  We
also note that in a trapping potential, $k$ is modified due to the
zero-point energy of the trap ($k_{ZP}$) and the effective thermal
wavevector for collisions is given by $k_T=\sqrt{(k^2+
k_{ZP}^2)/2}$. For our trap, $k_{ZP}\sim$3.5 $\mu$K, and $p$-wave
collisions are still suppressed. The observed density shift scales
as $G^{(2)}_{12}a_{ge}$, and for our typical temperatures we find values of
$G^{(2)}_{12}$ between 0.03 and 0.15, whereas the $p$-wave scattering
length is expected to be $\sim$1 \% of $a_{ge}$. Hence,
inhomogeneity-induced $s$-wave collisions dominate. In the the unitarity limit where $k_T|a_{ge}|> 1$ ($a_{ge}$ is the zero-temperature scattering length), the effective scattering length is 1/$k_T$. For our lattice trap parameters and temperature range of
1-3 $\mu$K, this length is on the order of -300 $a_0$, which is consistent in sign and magnitude with our observed frequency shifts, along with the values and uncertainties of $G^{(2)}_{12}$ and $\rho$.

To provide further evidence to exclude $p$-wave contributions, we
vary the inhomogeneity by misalignment of the spectroscopy probe
beam under a fixed $T$. This also helps rule out
$\bar{n}_{x,y,z}$ - dependent residual ac Stark shift of the trap.
Typically the probe beam is coaligned with the lattice to minimize
motional effects. However, by increasing the misalignment
($\Delta\theta$), we can also increase $\Delta\Omega$. Fig. 3A and B
show Rabi oscillations for two different probe beam misalignments at
$T$ = 1 $\mu$K (triangles and open squares) and 3 $\mu$K (circles
and open diamonds), respectively. Fig. 3C displays the measured
density shift as a function of
($\Delta\Omega$/$\overline{\Omega}$) due to probe misalignment.  For $T$ = 1 $\mu$K, the shift
becomes larger with increased $\Delta\Omega$/$\overline{\Omega}$.
When $\Delta\Omega$/$\overline{\Omega}$ increases further, the 3 $\mu$K data indicate that the density shift becomes
smaller. This behavior is reproduced by the theoretical curves shown
in Fig. 3C, and is illustrated in Fig. 3D. Consider two different
$\Delta\Omega$/$\overline{\Omega}$, both with an average excitation
fraction of 0.3. In the first case, for small misalignment, we find
a spread in the excitation fraction of $\pm$0.2; there is an
inhomogeneity allowing collisions to occur and we measure a small
density shift. In the second case, with further misalignment the
spread in the excitation fraction increases to $\pm$0.4; there is
now a larger spread in the Rabi frequencies, and collisions still
occur. However, we now have atoms with an excitation fraction both
above and below 50 \% where the shift crosses zero. Hence, the
collisions of atoms with excitations between 0.3$-$0.7 will average
to zero (this is consistent with the density shift going to zero at
50 \% excitation, regardless of the inhomogeneity), and the final
collision shift is due only to atoms with excitation fractions
between 0 and 0.3. The measured shift for the larger misalignment is
therefore smaller.

Combining the measurements shown in Figs. 2 and 3 makes it clear
that the observed density-dependent shifts arise  from the change of
the quantum statistics $G^{(2)}$ caused by the inhomogeneous
measurement process. Of course, the inhomogeneous effect can be
suppressed by decreasing the sample temperature and increasing the
transverse confinement, or going to higher dimension traps. Importantly, for
clock operations we have identified that near a 50 \% excitation
fraction the density shift goes to zero. Using these measurements we
can now reduce the uncertainty of the collision shifts for clock
operation\cite{Ludlow08} to 5$\times10^{-17}$. This time dependent
variation in quantum statistics will also apply to
boson-based clocks, where the original $G^{(2)}$ = 2 will decrease
to a value between 1 and 2.



\bibliography{densitypaper}

\begin{thebibliography}{10}
\bibitem{MITnote}
To mitigate the effects of inhomogeneous line broadening due to the density
  distribution of trapped atoms, Ketterle and
  co-workers\cite{Zwierlein03,Gupta03} spatially selected only a small central
  region when measuring the excitation fraction. By spatially selecting only a
  small fraction of trapped atoms where the trapping potential is more
  homogenous, the MIT group minimized their sensitivity to the inhomogeneous
  evolution described in this experiemnt.
\bibitem{SOM}
Materials and methods are detailed in supporting online material available at
  Science Online.
\bibitem{Thomsen08}
By analyzing the spectral components in sideband spectroscopy, the longitudinal
  temperature can be accurately determined. Extracting the transverse
  temperature is more complicated, however we have confirmed using
  time-of-flight analysis that the transverse and longitudinal temperatures are
  identical both before and after cooling (heating).
\bibitem{JulienneSP}
We have calculated the phase shifts, and corresponding lengths, using a model
  S+P potential with variable short-range shapes to change the scattering
  length over its full range. The short-range shape parameter varies so as to
  change the threshold phase and scattering length, corresponding approximately
  to changing the number of bound states in the potential by one. This
  represents the possible ranges of variation of any Sr van der Waals
  potential.
\end{thebibliography}

\bibliographystyle{Science}


\begin{scilastnote}
\item  We greatly appreciate technical contributions of T. Zelevinsky and insightful discussions with K. Gibble,
W. Ketterle, M. Zwierlein, and E. Cornell. We acknowledge funding
support from NIST, NSF, ONR, and DARPA. G. Campbell and A. D. Ludlow
are supported by National Research Council postdoctoral fellowships.
A.D.L.'s present address is NIST Time and Frequency Division. J. W.
Thomsen is a JILA visiting fellow, with a permanent address: The
Niels Bohr Institute, Universitetsparken 5, 2100 Copenhagen,
Denmark. The current address for P. Naidon is ERATO Macroscopic
Quantum Project, JST, Tokyo, 113-0033 Japan.

\end{scilastnote}


\clearpage

\begin{figure}
\includegraphics[hiresbb=true, width=7.5 cm]{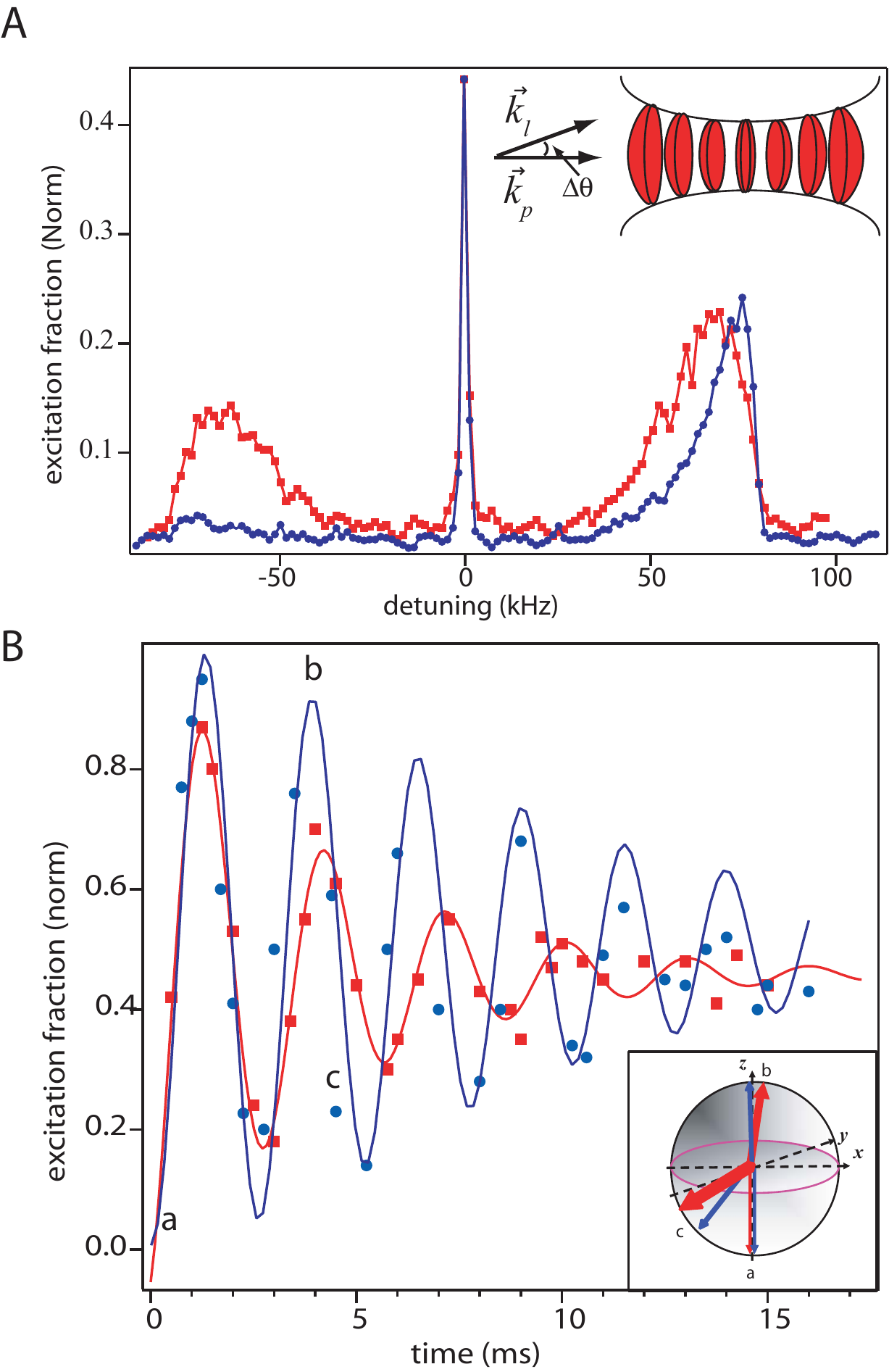}
\caption{(A) Sideband excitation spectra for $T$ = 1 $\mu$K (blue
circles) and 3 $\mu$K (red squares). The spectra are obtained in the
resolved sideband limit and have three dominant features, the narrow
carrier transition and broad red (blue) motional sidebands that are
excited when an atom is transferred to a lower (higher) motional
state during the transition. As the temperature of the sample is
lowered, the atoms primarily occupy the ground-state and the red
sideband is suppressed. The temperature of the atomic ensemble can
be extracted from a fit of the sidebands\cite{Thomsen08}.  The inset
shows the lattice geometry and excitation scheme. The probe beam and
lattice are coaligned and copolarized, minimizing the relative
spread between $\vec{k_l}$ and $\vec{k_p}$. However, even under the best effort, a small angle $\Delta\theta$
between the probe and lattice beams may persist due to aberrations and misalignment. (B) Rabi oscillations for
temperatures of 1 $\mu$K (blue circles) and 3 $\mu$K (red squares).
For higher temperatures, more motional states are occupied. This
leads to a larger spread in the Rabi frequencies and faster
dephasing of the excitation between atoms. By fitting the decay of
Rabi oscillations we can determine the degree of excitation
inhomogeneity. The inset illustrates the dephasing process using
rotations on the Bloch sphere. At time a, before the excitation, the
atoms are in a pure state. At time b, the atoms have undergone two
oscillations. For the red curve the temperature is hotter and there
is a larger spread in Rabi frequencies. This is indicated by the
increased width of the Bloch vector and dephasing of the observed
oscillations. At time c, the effect is even more pronounced.}
\end{figure}

\begin{figure}
\includegraphics[hiresbb=true]{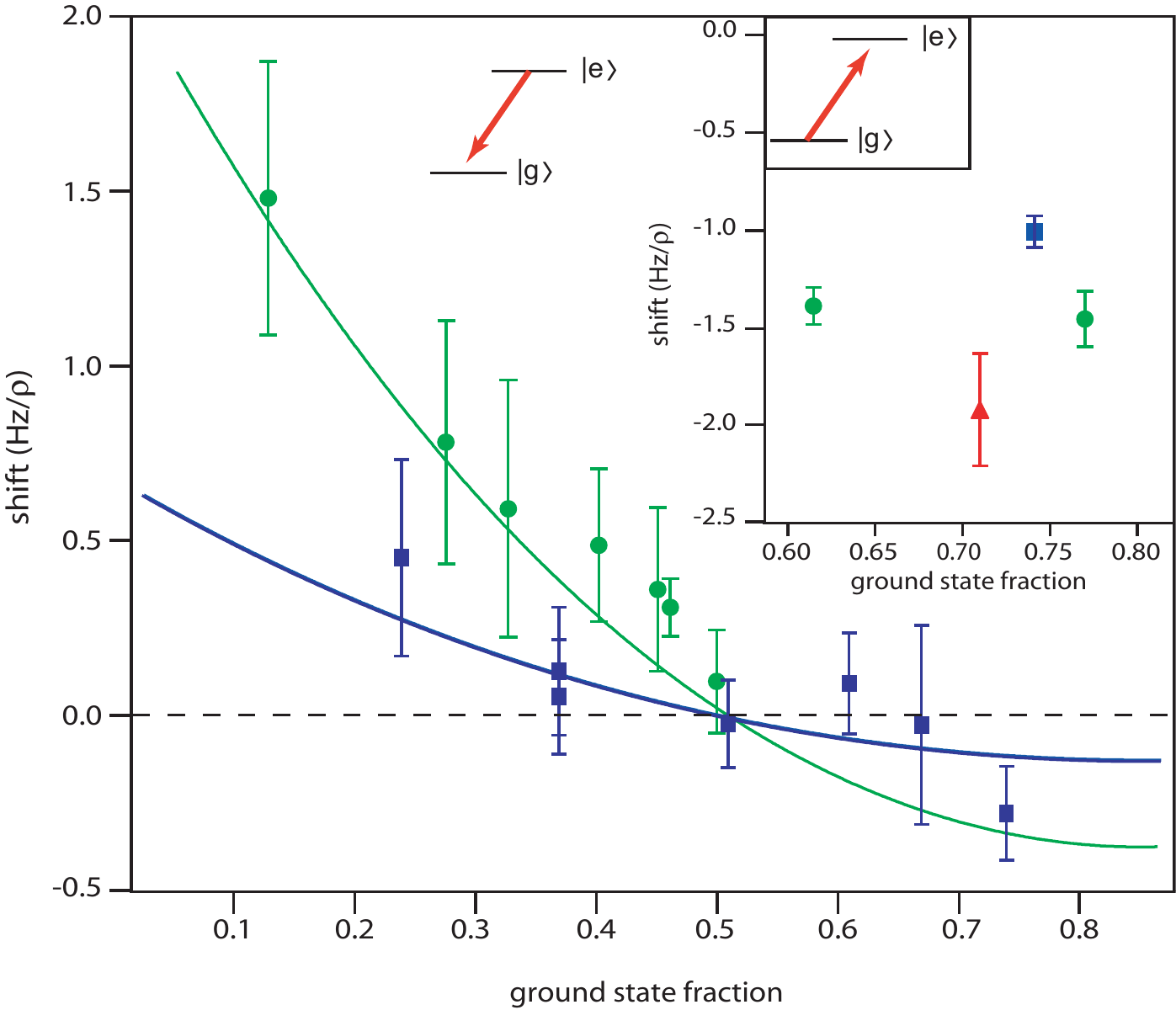}
\caption{Measured density-dependent frequency shift as a function of
the final excitation fraction and temperature. Atoms are initially
spin polarized and transferred to $^3$P$_0$ $(|e\rangle)$ before the
spectroscopy pulse is applied. The squares (circles) show the
measured shift for $T$ = 1 (3) $\mu$K. The lines show the calculated
shifts using the two-atom model with only a single scaling factor.
Near $\sim$50 \% the shift goes through zero. In the inset, the
measured shift is shown for atoms excited from the $^1$S$_0$
$(|g\rangle)$ state for $T$ = 1 $\mu$K (squares), 3 $\mu$K
(circles), and 5 $\mu$K (triangles). However, the magnitude in this
case could be influenced by imperfect spin polarizations.  For both
plots, as the temperature is decreased, the inhomogeneity also
decreases, leading to a smaller collision shift. $\rho$ is the
atomic density of $10^{11}$/cm$^3$. The density dependent shift for
each excitation fraction is determined using an interleaved scheme
where the density is varied every 100 s. Pairs of such data are then
used to determine the frequency shift. Typical data sets include 20
-- 30 pairs of density comparison, with the error bars indicating
the standard error.}
\end{figure}

\begin{figure}
\includegraphics[hiresbb=true, width=14 cm]{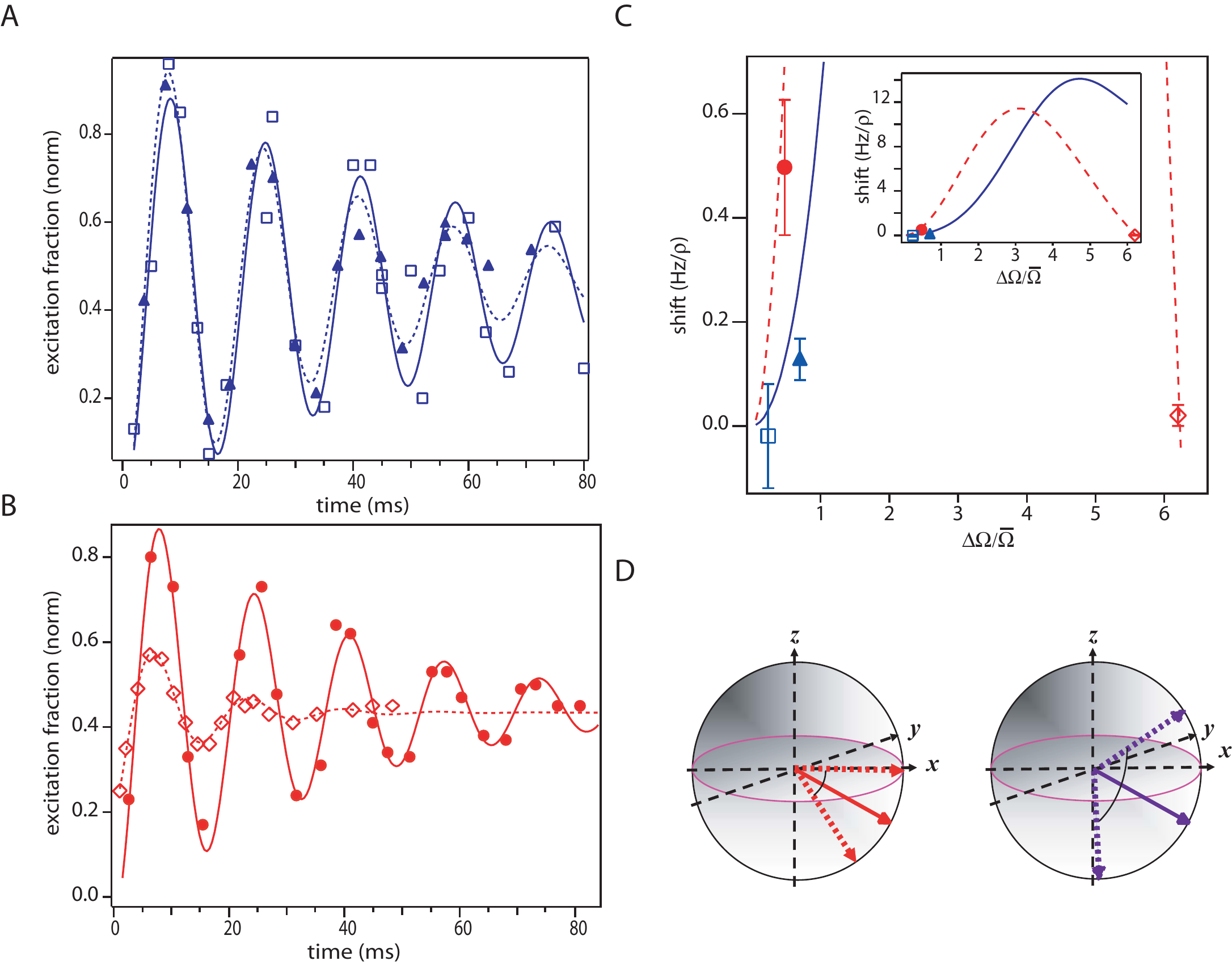}
\caption{Effect of probe misalignment on the density-dependent
shift. (A) Rabi oscillations are shown for two different values of
$\Delta\theta$ at $T$ = 1 $\mu$K. The open squares show oscillations
when the probe is aligned similar to that of Fig. 1 and 2. The solid triangles
show a faster dephasing when the probe beam misalignment is
increased further by 5 mrad. (B) Rabi oscillations for
$T$ = 3 $\mu$K. The circles show oscillations when
the probe beam is aligned similar to that of Fig. 1 and 2, and the diamonds when the
misaligment is increased further by 35 mrad. (C) The
density shift measured for each misalignment shown in (A) and (B).
>From $\Delta\theta$ and $T$, the spread in Rabi frequency
$\Delta\Omega$ is calculated. The lines show the expected shift as a
function of $\Delta\Omega$ for $T$ = 1 $\mu$K (solid line) and 3
$\mu$K (dashed line). The inset shows a zoomed-out plot. (D) For
large misalignments, we observe a smaller density shift. This is
described using the rotation on the Bloch sphere. As an example, two
different values of $\Delta\Omega$ are shown. On each sphere, the
average excitation fraction is shown with a solid line, and the
spread is indicated by the dotted lines. For small misalignments, we
have a small spread in Rabi frequencies. As the misalignment
increases, the spread crosses the equatorial plane of the Bloch
sphere. At 50 \%, the sign of the density shift changes, and
therefore the portion of the spread centered around this plane
averages to zero. The measured density shift is then reduced. }
\end{figure}

\end{document}



\baselineskip24pt


\maketitle

\section*{Materials and methods}

To first cool and trap the atoms, $^{87}$Sr is loaded into a
magneto-optical trap (MOT) from a slowed atomic beam. Fig. S1a shows
a diagram of relevant energy levels used in cooling, trapping and
detecting $^{87}$Sr. The MOT is operated on the strong
$^1$S$_0-^1$P$_1$ transition, and cools the atoms to millikelvin
temperatures. A second MOT stage operated on the narrow
$^1$S$_0-^3$P$_1$ transition is then used to cool the atoms further
to microkelvin temperatures. This second stage MOT uses
dual-frequency narrow line cooling\cite{Loftus2,Campbell08}. During
the MOT phase the one-dimensional optical lattice is superimposed on
the atoms in the vertical direction. After the second stage MOT, the
MOT cooling beams and magnetic gradient are turned off, leaving
$\sim$2000 atoms trapped in the optical lattice. After the lasers
and magnetic fields are turned off, the temperature of atoms
remaining in the lattice is $\sim$ 3 $\mu$K.

To vary the temperature of the atoms, additional cooling or heating
is performed on the lattice-trapped atoms.  The cooling is performed
in three dimensions, with Doppler cooling applied along the
transverse direction of the lattice, and sideband cooling (heating)
along the longitudinal axis.  The laser beams used for cooling
(heating) are red (blue) detuned from the $^1S_0$(F = 9/2)$-^3P_1$(F
= 11/2) transition. $\vec{F} = \vec{I} + \vec{J}$ is the total
angular momentum, with $\vec{I}$ the nuclear spin and $\vec{J}$ the
total electron angular momentum. The cooling (heating) beams are
applied first along the transverse direction for 20 ms, followed by
30 ms along the longitudinal direction. Simultaneous with the
sideband cooling (heating) the atoms are spin-polarized in a weak
magnetic bias field ($\sim$ 30 mG) using an optical pumping beam
resonant with the $^1$S$_0$(F = 9/2)$-^3$P$_1$(F = 9/2) transition.
By choosing the correct polarization for the optical pumping beam,
atoms are polarized into either the $m_F$ = +9/2 ($m_F$ = magnetic
quantum number) or $m_F$ = -9/2 ground state. To measure the final
temperature of the atoms, two different techniques are used. From
time-of-flight analysis the temperature of both the longitudinal and
transverse directions can be directly measured. By examining the
spectral components of sideband spectroscopy as shown in Fig. 1A,
the longitudinal temperature can also be accurately extracted. The
total atom number is determined by fluorescence measurements. The
average density is then calculated using the trap volume. The trap
volume is determined using our measured trap frequencies and
temperatures and assumes a Maxwell-Boltzmann distribution.

Orienting the lattice in the vertical direction breaks the energy
degeneracy between lattice sites, strongly prohibiting inter-site
tunneling\cite{Lemonde05}. This ensures that collisions occur only
between atoms in the same lattice site.  The lattice is aligned with
a slight angle with respect to gravity, to accommodate the vertical
MOT beams. The probe beam used for spectroscopy is aligned collinear
with the optical lattice, and the polarization of the beam is also
parallel to that of the lattice. To minimize inhomogeneity due to
probe beam focusing, the waist of the probe beam is $\sim$5 times
larger than the waist of the lattice focus.

\section*{Supporting text}
For a given atom in the electronic ground state $^1$S$_0$ and
motional state ($n_x$, $n_y$, $n_z$), the excitation probability
during the spectroscopy pulse is
\begin{equation}
p_e(\Delta,t)=\frac{\Omega_{\vec{n}}^2}{\Omega_{\vec{n}}^2+\Delta^2}\sin^2\left(\frac{
t\sqrt{\Omega^2_{\vec{n}}+\Delta^2}}{2}\right),
\end{equation}
where $\Delta$ is the detuning of the clock laser from resonance,
$t$ is the length of the spectroscopy pulse, and the modified Rabi
frequency is given by\cite{Wineland79}:
\begin{equation}
\label{eq:omega} \Omega_{\vec{n}}=\Omega_{n_x,n_y,n_z}=\Omega_0
e^{(-\eta_x^2-\eta_y^2-\eta_z^2)/2}L_{n_x}(\eta_x^2)L_{n_y}(\eta_y^2)L_{n_y}(\eta_y^2).
\end{equation}
Here $\Omega_0$ is the bare Rabi frequency, $L_{n_{x,y,z}}$ are Laguerre
polynomials, and $\eta_{x,y,z}$ are the Lamb-Dicke parameters for the
transverse and longitudinal directions (assuming the projection of the misalignment on the transverse plane bisects x and y axes):
\begin{align}
\label{eq:eta}
\eta_x&=\eta_y=\frac{\sin(\Delta\theta)}{\sqrt{2}\lambda_L}\sqrt{\frac{h}{2
m\nu_x}} \notag\\
\eta_z&=\frac{1}{\lambda_L}\sqrt{\frac{h}{2 m\nu_z}}.
\end{align}

To calculate the density shift, we return to our two-atom model. As
discussed in the main text we can explicitly calculate
$G^{(2)}_{12}$ from the anti-symmetrized overlap of the two
wavefunctions. The antisymmetrized wavefunction $|\Psi_{AS}\rangle$
is given by
\begin{align}
|\Psi_{AS}\rangle&=\frac{1}{\sqrt{2}}\left(\psi_1\psi_2-\psi_2\psi_1\right)
\\
&=\frac{1}{\sqrt{2}}\left((\alpha|g\rangle+\beta|e\rangle)(\gamma|g\rangle+\delta|e\rangle)-(\gamma|g\rangle+\delta|e\rangle)
(\alpha|g\rangle+\beta|e\rangle)\right)
\end{align}
and
\begin{align}
G^{(2)}_{12}(\alpha(t),\beta(t),\gamma(t),\delta(t))&=\langle\Psi_{AS}|\Psi_{AS}\rangle
\\ &=1-|\alpha\gamma^*+\beta\delta^*|^2.
\end{align}
 Each atom has a slightly different
$\Omega_{\vec{n}}$, as given in Eq.~\ref{eq:omega}, and
\begin{align}
\label{eq:alpha}
\alpha(t)&=i\frac{\Omega_1}{\Omega_{R1}}\sin\frac{\Omega_{R1}t}{2} \notag\\
\beta(t)&=\cos\frac{\Omega_{R1}t}{2}+i\frac{\Delta}{\Omega_{R1}}\sin\frac{\Omega_{R1}t}{2}
\notag
\\
\gamma(t)&=i\frac{\Omega_2}{\Omega_{R2}}\sin\frac{\Omega_{R2}t}{2}\notag
\\
\delta(t)&=\cos\frac{\Omega_{R2}t}{2}+i\frac{\Delta}{\Omega_{R2}}\sin\frac{\Omega_{R2}t}{2},
\end{align}
where $\Omega_{R1,2}=\sqrt{{\Omega_{1,2}}^2+\Delta^2}$ is the generalized
Rabi frequency. The inhomogeneity is a consequence of the spread in
Rabi frequencies $\Omega_{n_x,n_y,n_z}$, due to the excitation of
different motional states. We can characterize this by defining an
average Rabi frequency $\overline{\Omega}$ and its RMS spread
$\Delta\Omega$,
\begin{align}
\overline{\Omega}&=\sum_{n_x}\sum_{n_y}\sum_{n_z}q_x(n_x)q_y(n_y)q_z(n_z)\Omega_{n_x,n_y,n_z}
\\ \Delta\Omega^2&=\overline{\Omega^2}-{\overline{\Omega}}^2.
\end{align}
Here $q_{x,y,z}(n_{x,y,z})$ is the normalized Boltzmann factor.
The inhomogeneity in the Rabi frequencies ($\Delta\Omega$) is
affected by $T$ through the Boltzmann factor and by $\Delta\theta$
through the probe misalignment. To approximate the average density
shift for the entire atomic ensemble, we set $\Omega_1 =
\overline{\Omega}+\Delta\Omega$ and $\Omega_2 =
\overline{\Omega}-\Delta\Omega$ in Eq.~\ref{eq:alpha} for our
two-atom model. At a time $t$ during the spectroscopy pulse, the
atoms experience an ensemble-averaged shift:
\begin{equation}
\Delta\nu(t)=\frac{2\hbar
a_{ge}}{m}G^{(2)}_{12}(\overline{\Omega}+\Delta\Omega,\overline{\Omega}-\Delta\Omega)(\rho_g-\rho_e).
\end{equation}
This shift evolves during the spectroscopy pulse, and for the final
density shift we time-average $G^{(2)}_{12}$ over the total pulse
length $t_F$ as
\begin{equation}
\label{eq:finalshift}
\Delta\nu_{ave}=\frac{1}{t_F}\int_0^{t_F}\Delta\nu(t)dt.
\end{equation}
\begin{figure}
\includegraphics[hiresbb=true, width=8 cm]{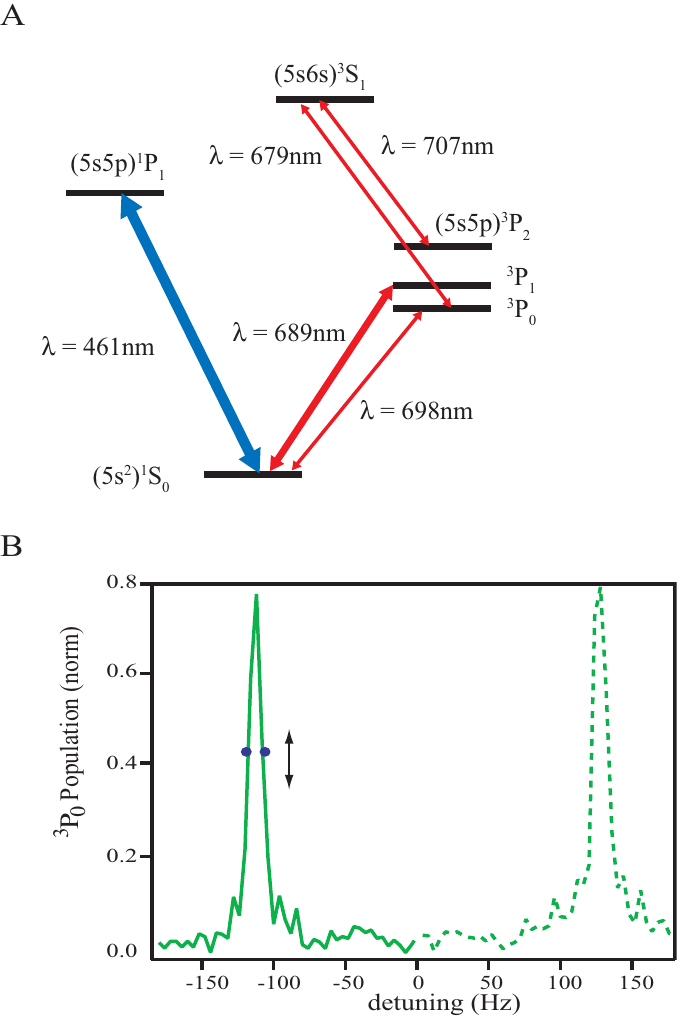}
\caption{Experimental setup for the $^{87}$Sr optical clock. (a)
Relevant energy levels for laser cooling and trapping are shown. The
first MOT stage is operated on the strong transition at 461 nm.
During this MOT stage atoms that are lost to the $^3$P$_2$ and
$^3$P$_0$ states are pumped back to the $^1$S$_0$ state using two
repump beams at 679 nm and 707 nm. The second MOT is operated on the
narrow 689 nm transition, and the wavelength of the clock laser used
for spectroscopy is 698 nm. After the experiment is performed, atoms
in the $^1$S$_0$ state are detected by measuring fluorescence on the
strong $^1$S$_0-^1$P$_1$ transition. Atoms that have been excited to
$^3$P$_0$ by the clock laser pulse are transferred back to the
$^1$S$_0$ using the two repumping beams, and are again measured by
fluorescence. (b) Two sample spectra. Here, $^{87}$Sr atoms are
successively polarized into either the $m_F$ = 9/2 or -9/2 state.
The separation between the two spectroscopy lines is $\sim$250 Hz.
During the actual experiment, we polarize the atoms to each pure
state successively, and lock near the full-width at half-maximum of
the peak (indicated by dots). By moving these lock points the
desired excitation fraction is selected.}
\end{figure}

\bibliography{densitypaper}

\bibliographystyle{Science}